# Deciphering the Role of Acetate in Metabolic Adaptation and Osimertinib Resistance in Non-Small Cell Lung Cancer

Giorgia Maroni (1), Eva Cabrera San Millan (1), Beatrice Campanella (2), Massimo Onor (2), Giovanni Cercignani (3), Beatrice Muscatello (4), Giulia Braccini (1), Raffaella Mercatelli (1), Alice Chiodi (1), Ettore Mosca (1), Elena Levantini (1)*#, Emilia Bramanti (2) *# (*co-last; #correspondance)

- *(1) CNR-Consiglio Nazionale delle Ricerche, Istituto di Tecnologie Biomediche (ITB-CNR), SS di Pisa, Area della Ricerca di Pisa, Italy*
- *(2) CNR-Consiglio Nazionale delle Ricerche, Istituto di Chimica dei Composti OrganoMetallici (ICCOM-CNR), SS di Pisa, Area della Ricerca di Pisa, Italy*
- *(3) Scuola Normale di Pisa, Class of Sciences, Pisa, Italy*
- *(4) Dipartimento di Farmacia, Università di Pisa, Pisa, Italy; Centro per l'Integrazione della Strumentazione dell'Università di Pisa (CISUP), Pisa, Italy*

**Abstract**

**Aims**. Resistance to targeted therapies remains a major challenge in EGFR-mutant non-small cell lung cancer (NSCLC). Here, we describe a novel metabolic adaptation in osimertinib-resistant cells characterized by elevated acetate levels and activation of an unconventional pyruvate-acetaldehyde-acetate (PAA) shunt. **Methods.** Integrated transcriptomic, exometabolomic, and functional analyses reveal suppression of canonical metabolic pathways and upregulation of ALDH2 and ALDH7A1, that mediate the $NADP^+$-dependent oxidation of acetaldehyde to acetate, generating NADPH. **Results.** This shift generates reducing power essential for biosynthesis and redox balance under conditions of oxidative pentose phosphate inhibition. These metabolic changes promote endurance in resistant cells and rewire the interplay between glycolysis, the pentose phosphate pathway, and the tricarboxylic acid cycle, offering a *de novo* bypass for anaplerosis and bioenergetics. Systematic metabolite profiling revealed distinct transcriptomic and metabolic signatures distinguishing resistant from drug sensitive parental cells.
**Conclusions.** Together, these findings depict a unique, resistance-driven adaptive metabolic shift and uncover potential therapeutic vulnerabilities in osimertinib-resistant NSCLC.

**Keywords:** EGFR-mutant NSCLC; osimertinib resistance; mitochondria; metabolic adaptation; acetate metabolism; pyruvate-acetaldehyde-acetate shunt; NADPH

1. **Introduction**

Mitochondrial oxidative phosphorylation (OXPHOS), localized to the inner mitochondrial membrane, is the primary source of ATP in most eukaryotic cells. OXPHOS couples electron flow along the respiratory chain [electron transport chain (ETC)] with ATP synthesis via ATP synthase, yielding ~30-34 ATPs per glucose molecule. Despite its high efficiency, OXPHOS is relatively slow and inherently generates reactive oxygen species (ROS) as byproducts, which can trigger oxidative damage to cellular components, thus promoting apoptosis.
In contrast, glycolysis is a faster, though less efficient process, producing only 2 ATPs per glucose molecule, and lower levels of ROS, thereby supporting cell survival under stress (Bose et al., 2019) (Gatenby and Gillies, 2004). In addition, its end-product, pyruvate, scavenges mitochondrial ROS, offering additional anti-apoptotic benefits (Fabrizio Marcucci and Rumio, 2021). The metabolic reprogramming from OXPHOS to glycolysis, known as the Warburg effect, is a hallmark of cancer (Diaz-Ruiz et al., 2011; Sun et al., 2018) and is increasingly implicated in therapeutic resistance (F Marcucci and Rumio, 2021; Sun et al., 2018), including resistance to EGFR tyrosine kinase inhibitors (TKIs) (Maroni et al., 2025).
When OXPHOS is compromised, cancer cells exhibit remarkable metabolic plasticity, by reprogramming energy pathways to sustain bioenergetics and redox balance under drug pressure.
Recent studies have uncovered complex metabolic mechanisms supporting resistance in NSCLC. Ali and Levantini et al. identified a mutant-EGFR/FASN (Fatty Acid Synthase) axis specific to gefitinib-resistant cells, in which palmitoylated nuclear EGFR contributes to acquired resistance (Ali et al., 2018). Additional adaptations include elevated cholesterol levels and upregulation of Caveolin-1 (Cav1), promoting GLUT3-mediated glucose uptake, selectively enhancing TKI-resistant cell survival, which can be targeted by statins to restore sensitivity (Ali et al., 2019). Mutant EGFR can also disrupt microtubule organization, impairing lysosomal trafficking and degradation of phosphorylated proteins that sustain oncogenic signaling; this mechanism can be pharmacologically targeted using microtubule stabilizers and lysosomal inhibitors (Chin et al., 2020). Together, these findings highlight the complexity of adaptive responses in TKI-resistant cells, involving coordinated genetic, transcriptional, metabolic, and intracellular trafficking rewiring that extends far beyond canonical resistance mutations (Ali et al., 2019, 2018; Chin et al., 2020; Levantini et al., 2022).

To further capture these adaptations, integrative transcriptomic and untargeted metabolomic approaches have been applied across a wide range of cancer types, including lung (Shi et al., 2019). These assays adopted the National Cancer Institute (NCI)-60 cell line panels (Rushing, 2023) as well as other established cancer cell line models (Andrei et al., 2020; Manjunath et al., 2022), overall highlighting key metabolic pathways associated with drug resistance.

However, untargeted analyses may overlook unconventional or poorly annotated metabolites that fall outside standard annotations. Thus, integrating untargeted analyses with systematic, targeted quantification of specific metabolites, complemented by transcriptional and enzymatic profiling offers a more comprehensive view of tumor metabolism. This comprehensive approach can potentially uncover critical blind spots and reveal alternative metabolic nodes that drive drug resistance, which may otherwise remain hidden in conventional bioinformatic-driven analyses.

One such underexploited, yet highly informative strategy, is exometabolomic profiling, i.e., the targeted analysis of metabolites contained within the extracellular medium (ECM). As a dynamic interface between cells, the ECM reflects shifts in metabolic flux and offers real-time readouts of anabolic and catabolic activities (Meiser et al., 2018). This methodology efficiently identifies biomarkers linked to metabolic adaptations in cell culture (Meiser et al., 2018). Importantly, this minimally invasive method preserves analyte integrity and quantity by avoiding cell lysis and the matrix effects caused by extraction artifacts (Bramanti et al., 2019a; Campanella et al., 2021; Colombaioni et al., 2022, 2017). Using the cell culture *exometabolome*, previous studies have explored cancer metabolism and intercellular communications that promote tumor formation and progression (Cadenas-De Miguel et al., 2023; Meiser et al., 2018). Therefore, we sought to extend the same approach to the context of drug resistance.

Here, we employed high-resolution chromatographic platforms, including reversed-phase liquid chromatography (HPLC-DAD and LC-HRMS) and GC-MS, to systematically quantify ECM metabolites (Bramanti et al., 2019b; Campanella et al., 2021) in parental (Par) and osimertinib resistant (OsiR) H1975 NSCLC cells, harboring the common L858R/T790M EGFR mutations. Pathway enrichment analysis of the exometabolome data was applied to explore the metabolic pathways involved in the drug resistant mechanism. Interestingly, our targeted analyses revealed a remarkable accumulation of acetate, along with elevated pyruvate, acetaldehyde and lactate in resistant cells.

Previous studies suggest a *de novo* pathway for acetate production from pyruvate (Alderton, 2015; Bose et al., 2019; Busch, 1953; Emmelot and Bosch, 1955; Lakhter et al., 2016; Liu et al., 2018; Schug et al., 2016; Yoshii et al., 2015). Liu et al. demonstrated in several model cell lines that acetate can be chemically generated through a non-enzymatic reaction in which ROS directly target pyruvate via nucleophilic attack, converting it into acetate (Liu et al., 2018). In this reaction, pyruvate acts as a key ROS scavenger, a protective mechanism triggered under ETC dysfunction, which also preserves NADH for its essential role in reducing pyruvate to lactate (Kong and Chandel, 2018).

However, the majority (> 90%) of cellular acetate production appears to derive from a neomorphic enzyme activity of keto acid dehydrogenases (KDHs) (Liu et al., 2018) like pyruvate dehydrogenase E1 (PDHE1). These KDH enzymes, located in both the inner mitochondrial membrane and the nucleus, depend on multiple cofactors, including pyruvate, CoA, $NAD^+$, ROS, and thiamine/thiamine pyrophosphate (TPP) (Liu et al., 2018). PDHE1 canonically catalyzes the decarboxylation of pyruvic acid in a TPP- and magnesium-dependent reaction, generating acetyl-CoA. Liu et al. proposed instead a model in which imbalances in substrates and cofactor concentrations can modify PDH function to produce acetaldehyde and acetate instead of acetyl-CoA (Liu et al., 2018). Importantly, in metabolically hyperactive tumors, acetate is often released into the extracellular space (Liberti and Locasale, 2016), supporting metabolic symbiosis among neighboring cancer cells (Bose et al., 2019).

In light of these premises our findings are discussed to understand how the reorganization of central carbon metabolism in osimertinib-resistant NSCLC cells supports cancer cell survival, characterized by increased glycolysis, decreased oxidative pentose phosphate pathway (ox-PPP) activity, and activation of the alternative pyruvate-acetate-acetaldehyde pathway. This metabolic signature not only characterizes the resistant phenotype but also reveals a metabolic vulnerability in the acetate-NADPH axis, offering new therapeutic opportunities against refractory disease.

## 2. Materials and Methods

### 2.1. Chemicals and procedures

Sulfuric acid for HPLC analysis (30743, Honeywell Fluka, 95-97%) and methanol for RP-HPLC (34860, Merck, 99.9%) were used. Standard solutions for HPLC (TraceCERT®, 1000 mg/L in water) were purchased from Sigma-Aldrich (Milan, Italy). Stock solutions of analytes were prepared by dissolving weighed amounts of pure compounds in deionized water and stored at 4°C for up to one month. All sample and solution preparations/dilutions were performed gravimetrically using ultrapure water (MilliQ; 18.2 MΩ cm-1 at 25 °C, Millipore, Bedford, MA, USA).

### 2.2. Cell cultures

H1975 parental (Par) and Osimertinib-resistant (OsiR) cell lines were cultured for 72 h in RPMI 1640 medium supplemented with 10% fetal bovine serum (FBS, Sigma-Aldrich) at 37 °C in a humidified incubator at 5% $CO_2$. Cells were authenticated by DNA fingerprinting and tested negative for mycoplasma contamination.

### 2.3. Exometabolome analysis

Following 72 h incubation, the extracellular medium was collected from cultured cells and subjected to downstream analysis.

### 2.4. Pyruvate quantification by LC-HRMS

Pyruvate levels were quantified by LCMS. The analysis was carried out on an ultra-high-performance liquid chromatography (UHPLC; Vanquish Flex Binary pump) coupled to a diode array detector (DAD) and a high-resolution (HR) Q Exactive Plus mass spectrometer (MS), based on Orbitrap technology, equipped with a heated electrospray ionization (HESI) source (Thermo Fischer Scientific Inc., Bremen, Germany).

Chromatographic separation was achieved on a Zorbax® Eclipse Plus Phenyl-Hexyl column (Agilent, CA, United States; 4.6 x 250 mm, 5 μm). The flow rate was 0.8 mL/min, split 1:1 between the MS and DAD/UV detector. The column temperature was maintained at 40°C, and the injection volume was 2 μL.

Eluents consisted of MeOH/HCOOH 0.5% v/v (solvent B); $H_2O$/HCOOH 0.5% v/v (solvent A) (all solvents were of ultra-purity grade, ROMIL Ltd, Cambridge, GB).

The gradient was as follows: isocratic 100% A until 15 min, linear gradient from 0 to 80% (B) in the range 15-25 min, from 80% to 100% (B) in the range 25-35 min and isocratic 100%B for 2 min. The chromatographic run was complete in 44 min including re-equilibration in 100% solvent A. Ionization parameters were: nebulization voltage 3400 V (+) and 3200 V (-), capillary temperature 290°C, sheath gas ($N_2$) 24 (+) and 28 (-) arbitrary units, auxiliary gas ($N_2$) 5 (+) and 4 (-) arbitrary units, S-lens RF level 50. Scan mode: Full scan, Scan range: m/z 54-800. Resolution 70,000 at m/z 200. Polarity: positive and negative ionization mode.

Data were acquired and analyzed by Xcalibur 3.1 software (Thermo Fischer Scientific Inc., Bremen, Germany). LCMS files were processed using Xcalibur 4.1 (Thermo Fisher Scientific). Extracted ion chromatograms (EICs) were obtained using exact mass range calculations based on the elemental formula ($C_nH_nO_nP_aN_nS_n$ - H for negative ions; +H for positive ions) with ± 3 ppm ($\Delta M/M$) $*10^6$ of accuracy. Peaks were manually integrated, and results elaborated in Microsoft Excel.

*2.5. Lactate quantification by RP-HPLC-DAD*

Lactate concentrations in the extracellular medium were measured by reversed-phase liquid chromatography (RP-HPLC) equipped with diode array and fluorescence detectors (HPLC-DAD) (Campanella et al., 2020). RP-HPLC-DAD was used to avoid MS detector saturation due to high lactate concentration (mM range) in the extracellular compartment. Samples were diluted 1:5 in 5 mM sulfuric acid, filtered through 0.20 mm RC Mini-Uniprep filters (Agilent Technologies, Italy), and injected ($V_{inj}$ = 5 μL) into the HPLC system. The same column used for LC-HRMS was employed. Lactate was identified by comparing retention times and UV spectra with standards. Signals were manually integrated, and concentrations were determined from calibration curves of analytical standards.

*2.6. Acetate and acetaldehyde quantification by HS-GC-MS*

Acetate and acetaldehyde levels were quantified by headspace (HS) GC-MS. Acetate was determined after acidification with $H_2SO_4$ to convert it to volatile acetic acid via (HS) GC-MS.

For acetate analysis, 500 μL of sample or standard acetate (0, 50, 100, 250, 400 μg/mL; prepared gravimetrically from Certified Standard), 50 μL of $^2H_3$ acetate at 1000 μg/mL (Internal Standard), and 100 μL of 5 M $H_2SO_4$ were added to 10 mL HS vials and sealed immediately. Standards and samples were prepared in triplicate. Calibration curves were generated by plotting the area ratio (m/z 43 /area at m/z 46) versus the concentration ratio [acetate]/[acetate $^2H_3$] using Agilent MassHunter Quantitative Analysis Ver. 10.2.

A GC Agilent 6850 together with an Agilent single quadrupole MS 5975c, and equipped with an Agilent GC 80 CTC PAL ALS was employed. Separations for acetate were achieved using a J&W DB-WAX-UI capillary column (30 m, 250 μm internal diameter, 0.50 μm film thickness) with constant flow of 1.0 mL/min (average velocity: 36 cm/sec; incubation at 80 °C for 600 s; 1000 μL injection volume). Oven program was as follows: initial temperature 50° C, hold 5 min, ramp 10° C/min to 150° C hold for 2 min, ramp 20° C/min to 240 °C hold 8.5 min (total 30 min). Pulsed Splitless inlet mode; initial temperature 200° C, pressure 45.4 kPa, pulse pressure 120 kPa, pulse time 0.10 min, purge flow 200 mL/min, purge time 0.10 min, total flow 203.7 mL/min, gas type helium; transfer line temperature 250 °C; SIM acquisition mode. MS conditions were as follows: electron ionization (EI) mode at 70eV electron energy; Ion Source Temp. 250° C Quadrupole Temp. 150° C; Acquisition mode SIM; acquisition m/z 43, 46, 60, 63; Dwell Time 100 ms; quantification ions: 43 and 46; qualifier ions: 60 and 63. Acquisition Software: Agilent MSD ChemStation Ver. E.02.02.

For acetaldehyde analysis, 500 μL of sample or standard acetaldehyde (0, 50, 100, 250, 400 μg/mL; prepared gravimetrically from Certified Standard) were spiked with 50 μL of $^2H_3$ ethanol at 89.2 μg/mL (Internal Standard). Separations for acetaldehyde were achieved using the same column with a constant flow of 1.0 mL/min (average velocity: 36 cm/sec; incubation at 25° C for 600 s; 1000 μL injection volume). Oven program was as follows: initial temperature 25° C, hold 5 min, 25° C/min to 240° C hold 5.00 min. Quantification ions: 29.44 for acetaldehyde, 45 and 46 for ethanol, and 48 and 49 for $^2H_3$. Detailed description of protocols was as previously reported (Campanella et al., 2019).

*2.7. SDS-PAGE and Western Blotting*

For protein analysis, 15 μg of protein from each sample (Par and OsiR cells) was separated by SDS-PAGE on 10% gel and subsequently transferred to nitrocellulose membranes by the TransBlot Transfer System (Bio-Rad). Membranes were blocked for 1 hour at room temperature in either 5% non-fat dry milk or 5% bovine serum albumin (BSA) diluted in Tris-buffered saline with 0.1% Tween-20 (1X TBST). Primary antibodies (listed below) were incubated overnight at 4°C. After incubation, membranes were washed three times for 10 minutes each with 1X TBS-T at room temperature. Horseradish peroxidase (HRP)-conjugated secondary antibodies (listed below) were then applied for 1 hour at room temperature. Proteins bands were detected by chemiluminescence (Pierce ECL Western Blotting Substrate Cat #32106) and imaged using the Chemidoc Imaging System (Biorad). Membranes were subsequently stripped using Stripping Buffer Solution (Himedia, #ML163), as per the manufacturer's instructions, and re-probed with an anti-β-actin mouse antibody to verify equal protein loading. After three washes with 1X TBS-T, membranes were incubated with the appropriate anti-mouse secondary antibody. Primary antibodies were the following: anti-β-actin mouse antibody (Santa Cruz #C3022) (1/1000 in 5% non fat milk dilution); anti-PDHE1 rabbit antibody (Proteintech #18068-1-AP)(1/1000 in 5% non fat milk dilution); anti-PFKFB4 rabbit antibody (Proteintech #29902-1-AP)(1/2000 in 5% non fat milk dilution); anti-ACSS1 rabbit antibody (Thermofisher #109357) (1/1000 (in 5% non fat milk dilution). Secondary antibodies were the following: anti-mouse (Immunoreagents #NC27606) (1/2000 in 5% non fat milk dilution); anti-rabbit (Cell Signaling Technologies #7074) (1/2000 in 5% non fat milk dilution).

*2.8. Pathway analysis.* Pathway topology and biomarker analysis of discriminating metabolites was carried out by using MetaboAnalyst 6.0.(Pang et al., 2024b, 2024a)

3. **Results**

*3.1 Exometabolomic Profiling by LC-HRMS, LC-DAD and GC-MS: Metabolic Indicators in Osimertinib-resistant cells*

Bioinformatic analysis revealed substantial metabolic rewiring in OsiR cells, affecting core bioenergetic pathways, including glycolysis/gluconeogenesis, pyruvate metabolism, glyoxylate and dicarboxylate metabolism, as well as vitamin and amino acid pathways (**Figure 1** and **Table 1**). These findings were supported by transcriptomic and complementary biological assays. We previously demonstrated that OsiR cells undergo mitochondrial and metabolic remodeling, characterized by impaired OXPHOS, structural mitochondrial alterations, accumulation of mtDNA mutations, enhanced glycolysis, and activation of the pyruvate–acetaldehyde–acetate pathway (Maroni et al., 2025).

Here, we extend these observations through a targeted analysis of gene expression in the Maroni et al. dataset, identifying key enzymes that drive metabolic and mitochondrial adaptation. This analysis confirms a mechanistic link between metabolic rewiring and osimertinib resistance, providing a detailed map of the adaptations underlying drug tolerance and highlighting potential vulnerabilities. Altered mitochondrial function and morphology, impaired ETC activity, and compromised OXPHOS, likely due to mtDNA mutations and/or coordinated changes in gene expression, collectively contribute to the metabolic reprogramming observed in OsiR cells.

In the following, we discuss these metabolic adaptations in detail, emphasizing pathway-specific alterations and their potential contribution to drug resistance.

**Figure 1(A)** shows the pathway analysis of significantly altered exometabolites in OsiR *vs* Par cells, as determined using LC-HRMS, LC-DAD and GC-MS. Pathways were mapped via the KEGG database and ranked according to pathway impact and adjusted *p*-value ($-\log(p\ value)$), highlighting the most altered metabolic pathways distinguishing OsiR from Par cells. Comparable results were obtained through metabolite set enrichment analysis (MSEA) (**Figure 1B**), a metabolomics approach that evaluates the enrichment of predefined metabolic pathways to identify those most significantly perturbed (Pang et al., 2024b).

Pathway topology and biomarker analyses identified several pathways significantly dysregulated in OsiR cells, including glycolysis/gluconeogenesis, pyruvate metabolism, phenylalanine, tyrosine and tryptophan biosynthesis, glyoxylate and dicarboxylate metabolism, nicotinate and nicotinamide metabolism, and phenylalanine metabolism (**Table 1**). Four metabolites specific to glycolysis and related pathways (pyruvate, lactate, acetate, and acetaldehyde) were quantified in the exometabolome of Par and OsiR cells cultured for 72 h. Their concentrations (mean ± SD; n=3 biological replicates), normalized to cell numbers, are summarized in **Table 2**.

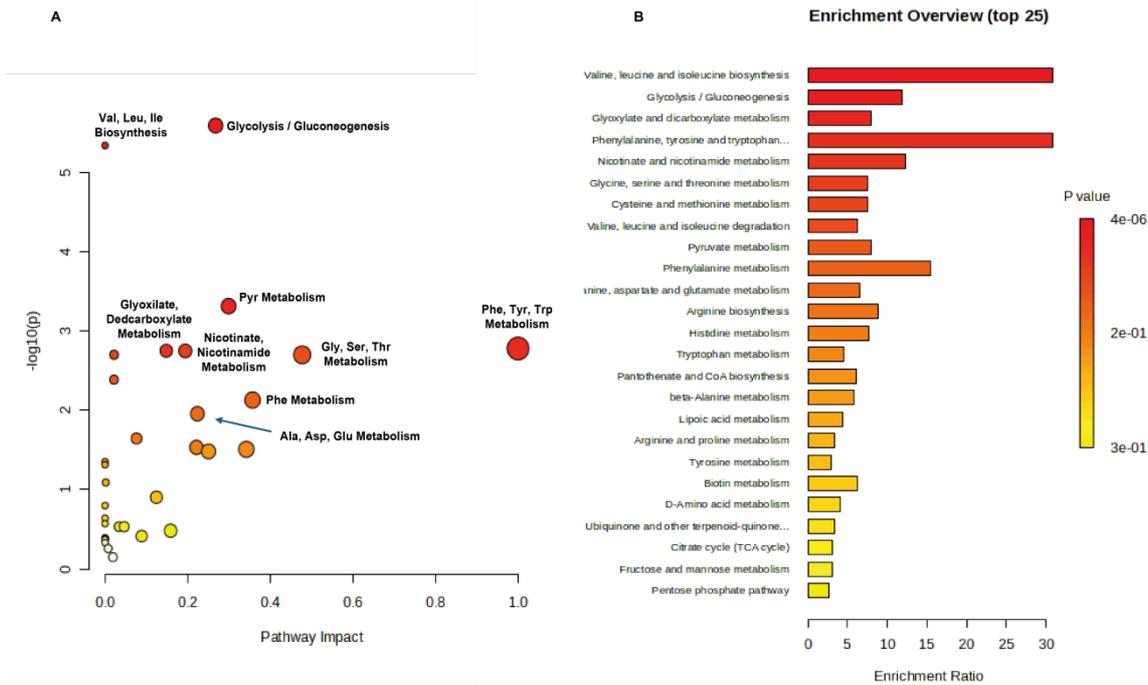

**Figure 1. (A)** Pathway analysis of significantly altered exometabolites in OsiR *vs* Par cells determined by LC-HRMS, LC-DAD and GC-MS. Pathways were mapped via the KEGG database and displayed according to pathway impact (dot size) and adjusted *p*-value (color intensity; darker = more significant). **(B)** Metabolite set enrichment analysis confirmed the most impacted pathways in OsiR vs Par cells. Statistical and topological analyses were performed using MetaboAnalyst 6.0(Pang et al., 2024b, 2024a). Pathway impact represents a combination of pathway enrichment and topology analysis (centrality). Circle size indicates pathway impact, while color intensity represents statistical significance (more intense red color corresponds to lower *p*-values, and higher significance).

**Table 1**. Top 10 significantly dysregulated metabolic pathways in OsiR cells.

|  | Total | Expected | Hits | Raw *p* | -log10(*p*) | Holm adjust | FDR | Impact |
|---|---|---|---|---|---|---|---|---|
| **Glycolysis / Gluconeogenesis** | 26 | 0.44571 | 6 | 2.5748E-06 | 5.5893 | 0.00020598 | 0.00018353 | 0.2677 |
| **Valine, leucine and isoleucine biosynthesis** | 8 | 0.13714 | 4 | 4.5883E-06 | 5.3383 | 0.00036248 | 0.00018353 | 0 |
| **Pyruvate metabolism** | 23 | 0.39429 | 4 | 0.00048638 | 3.313 | 0.037937 | 0.01297 | 0.29918 |
| **Phenylalanine, tyrosine and tryptophan biosynthesis** | 4 | 0.068571 | 2 | 0.0016632 | 2.779 | 0.12807 | 0.019978 | 1 |
| **Glyoxylate and dicarboxylate metabolism** | 32 | 0.54857 | 4 | 0.0017764 | 2.7505 | 0.13501 | 0.019978 | 0.14815 |
| **Nicotinate and nicotinamide metabolism** | 15 | 0.25714 | 3 | 0.0017838 | 2.7487 | 0.13501 | 0.019978 | 0.1943 |
| **Cysteine and methionine metabolism** | 33 | 0.56571 | 4 | 0.0019978 | 2.6995 | 0.14783 | 0.019978 | 0.02184 |
| **Glycine, serine and threonine metabolism** | 33 | 0.56571 | 4 | 0.0019978 | 2.6995 | 0.14783 | 0.019978 | 0.47737 |
| **Valine, leucine and isoleucine degradation** | 40 | 0.68571 | 4 | 0.0041085 | 2.3863 | 0.29581 | 0.03652 | 0.02168 |
| **Phenylalanine metabolism** | 8 | 0.13714 | 2 | 0.007439 | 2.1285 | 0.52817 | 0.059512 | 0.35714 |

**Table 2**. Concentration levels of pyruvate, lactate, acetate, and acetaldehyde in the ECM of Par and OsiR cell cultures. Measurements were performed on three biological replicates and normalized to cell numbers. LOD= limit of detection.

| Concentration | Method | Unit | mean ± SD (Par) | mean ± SD (OsiR) |
| --- | --- | --- | --- | --- |
| Pyruvate | (LC-HRMS) | (nmol/cell) | 0.022±0.004 | 0.113± 0.018 |
| Lactate | (LC-DAD) | (nmol/cell) | 0.429±0.021 | 0.606± 0.047 |
| Acetate | (HS-GCMS) | (nmol/cell) | <LOD | 1.144± 0.219 |
| Acetaldehyde | (HS-GCMS) | (pmol/cell) | 0.0019±0.0005 | 0.0050± 0.0009 |

Along with pyruvate, acetate, lactate and acetaldehyde, **Figure 2** shows also the bar plots of the peak areas/concentrations of glyceraldehyde-3-phosphate (G3P), ribose-5-phosphate, D-ribulose-5-phosphate and D-xylulose-5-phosphate, and 3-phosphoglycerate (3PG), by LC-HRMS, HPLC-DAD and HS-GCMS in Par and OsiR ECMs. LC-HRMS analysis evidenced a **decrease in G3P levels** in OsiR cells (**Figure 2C**) and an increase in ribose-5-phosphate (and its isomers), which are metabolites of the non-ox PPP, and 3PG (**Figure 2E**), which is a glycolytic metabolite.

To validate the metabolic reprogramming suggested by the exometabolomic profiling, we analyzed selected enzymes by Western Blot (**Figure 2H-J**).
Several key enzymes in glycolysis, the PPP, tricarboxylic acid (TCA) cycle, and aldehyde detoxification, including transketolase (TKT), pyruvate dehydrogenase complex (PDH), succinate dehydrogenase (SDH), aldolase A (ALDOA), aldehyde dehydrogenase 2 (ALDH2) and aldehyde dehydrogenase 7 family member A1 (ALDH7A1), were reported as upregulated in OsiR cells in our earlier study (Maroni et al., 2025). Here we assessed the increased expression of glycolytic regulator 6-Phosphofructo-2-Kinase/Fructose-2,6-Biphosphatase 4 (PFKFB4) (**Figure 2H**), and pyruvate dehydrogenase E1 mediating pyruvate decarboxylation (**Figure 2I**). Acyl-CoA synthetase short-chain family member 1 (ACSS1), which converts acetate to acetyl CoA (**Figure 2J**), was instead mildly upregulated. Taken as a whole, all three enzymes' activities were increased in OsiR cells, providing direct protein-level evidence supporting the exometabolomic findings. These results highlight enhanced glycolysis, pyruvate utilization, and acetate-driven acetyl-CoA production in OsiR metabolic reprogramming.

While transcriptomics shows that **TKT** is equally expressed in Par and OsiR cells, Western blot evidenced TKT overexpression in OsiR cells (Maroni et al., 2025). This correlates with the observed increase in ribose-5-phosphate, D-ribulose 5-phosphate, and D-Xylulose 5-phosphate in OsiR ECMs (**Figure 2D**).

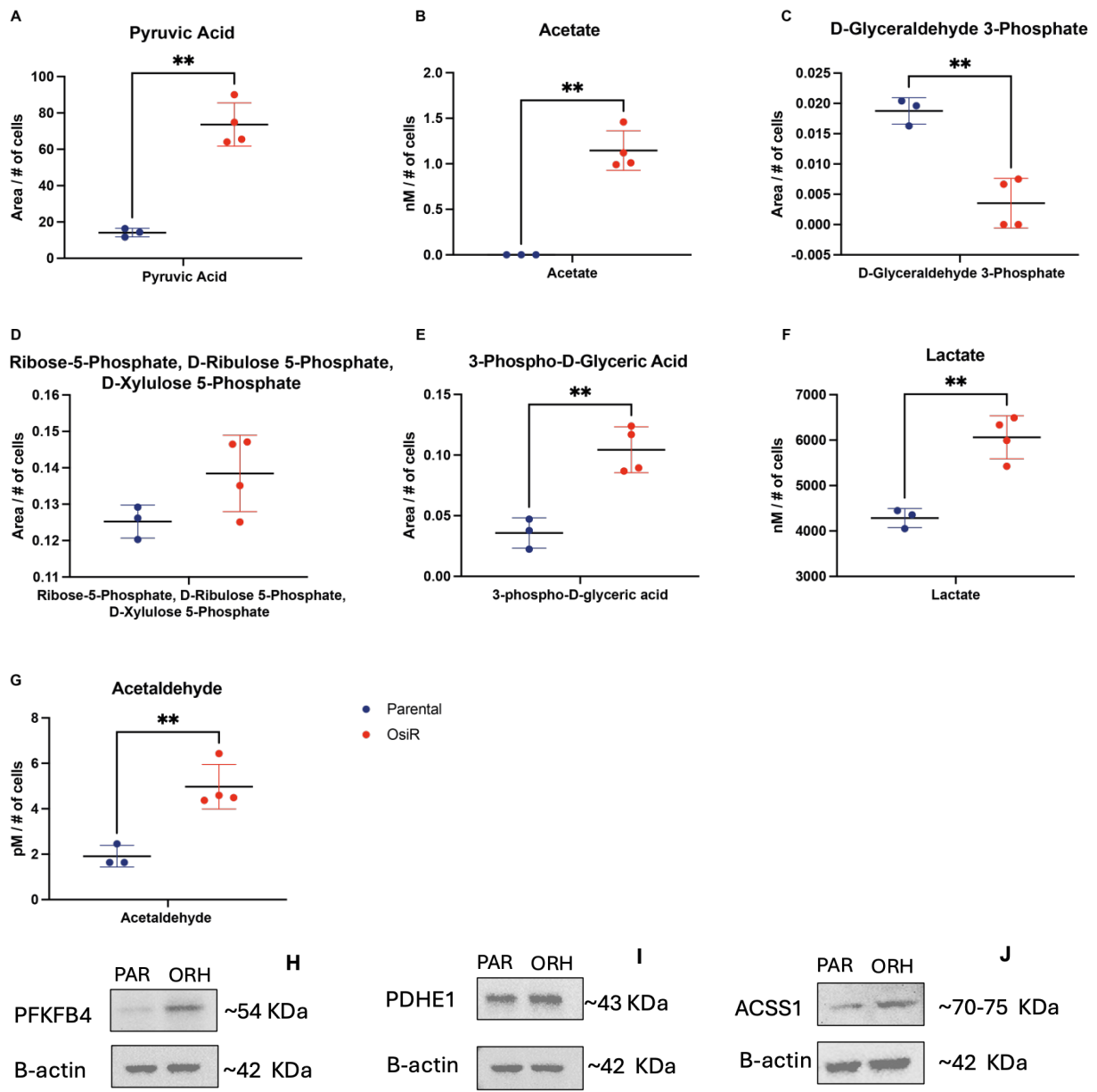

**Figure 2**. **(A-G)** Bar plots showing the peak areas (concentrations) of pyruvate, acetate, lactate, acetaldehyde, and other glycolysis/gluconeogenesis-associated metabolites measured by LC-HRMS, HPLC-DAD and HS-GCMS in Par (blue dots) and OsiR (red dots) ECMs. Statistical significance: $p< 0.001$ (**); $p<0.05$ (*). **(H-J)** Western blot analyses of Par and OsiR cells. Protein lysates were immunoblotted with an anti-PFKFB4, anti-PDHE1 and anti-ACSS1 antibody. Loading was assessed with an anti-β-actin antibody. Expected size is indicated in kDa.

3.2 *Transcriptomic Profiling and Gene Expression Changes in OsiR Cells*

RNAseq data showed modulation of key regulatory enzymes related to the glycolytic pathway, oxPPP, phosphofructokinase regulation, pyruvate metabolism, and TCA, resulting either up- or down- regulated in OsiR and Par cells, evidenced as heatmaps in **Figure 3**.

Hexokinase HK1 is considered the housekeeping isoform. In OsiR cells, **HK1 is upregulated** (**Figure 3**), while **HK2 is downregulated**, thereby ensuring glucose is committed toward glycolysis.

Once phosphorylated, glucose can enter either glycolysis or the PPP. In OsiR cells, transcription of both microsomal hexose-6-phosphate dehydrogenase (**H6PD)**, and cytosolic **G6PD** are downregulated, along with the 6-phosphogluconolactonase (**PGLS**) gene.

While the genes encoding ox-PPP enzymes (H6PD, G6PD and PGLS) are downregulated is OsiR cells, several key glycolytic enzymes, such as phosphofructokinase, aldolase A and enolase isoforms, are upregulated, consistent with enhanced glycolytic activity.

**Phosphofructokinase** (PFK1) catalyzes the first uniquely committed step of glycolysis, the irreversible conversion of fructose 6 phosphate (F6P) to fructose 1,6 biphosphate (F1,6BP), exclusively used for glycolysis. In OsiR cells, transcriptomic analysis revealed **upregulation of PFKM** and **downregulation of PFKL (Figure 3).** Moreover, **PFKFB3** and **PFKFB4** isoenzymes of 6-phosphofructo-2-kinase/fructose-2,6-bisphosphatase (F2,6BP) are upregulated in OsiR cells, controlling cytoplasmic F2,6BP levels and allosterically activating PFK1. Western Blot analysis confirmed the upregulation of PFKFB4 at the protein level (**Figure 2K**).

Supporting this glycolytic upregulation, key enzymes such as **aldolase A** and **enolase isoforms ENO1 and ENO2** are overexpressed in OsiR cells. **ALDOA** catalyzes the reversible conversion of F1,6PB to glyceraldehyde-3-phosphate (G3P) and dihydroxyacetone phosphate (DHAP), a critical glycolytic node, while **ENO1** and **ENO2** catalyze the reversible conversion of 2-phosphoglycerate (**2PG**) to phosphoenolpyruvate (PEP).

Thus, the glycolytic pathway produces pyruvate, which can serve as a substrate for different enzymatic reactions: conversion to lactate, as reflected by the upregulation of lactate dehydrogenases; or entry into the TCA cycle as Acetyl-CoA, supported by increased expression of pyruvate dehydrogenase (PDHB). Notably, pyruvate, upon dehydrogenation, can also contribute to acetaldehyde and acetate production, both increased in OsiR cells, as previously shown, consistent with the observed upregulation of acetaldehyde dehydrogenases (ALDH2, ALDH7A1). In contrast, no significant differences were detected in nicotinamide nucleotide transhydrogenase (**NNT**) gene expression between Par and OR cells (data not shown). NNT, an integral protein of the inner mitochondrial membrane, catalyzes the hydride transfer between NADH and $NADP^+$ coupled to proton translocation across the membrane, exploiting the mitochondrial proton gradient to produce high concentrations of NADPH, a critical reducing equivalent.

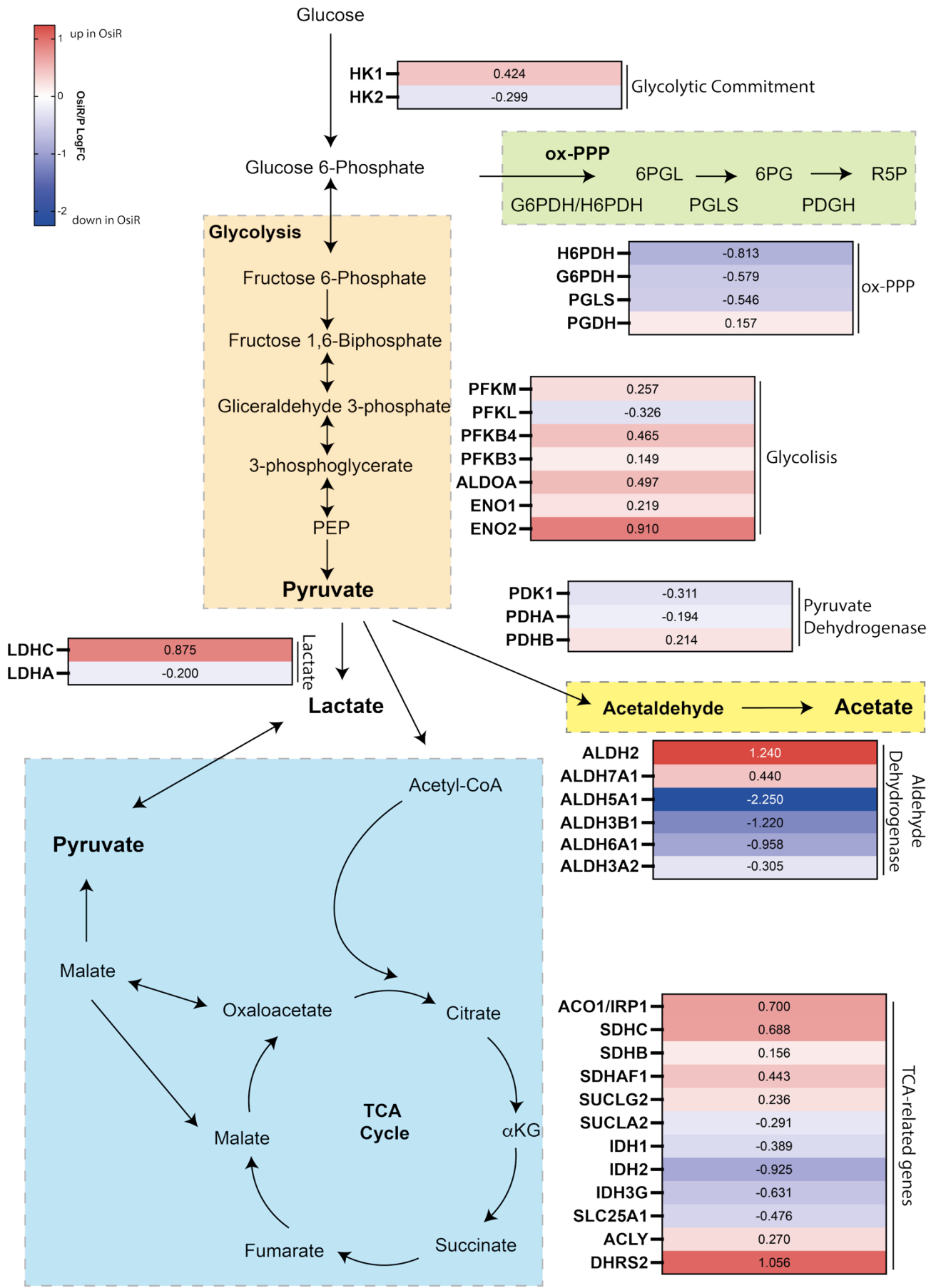

**Figure 3**. Heatmap of key regulatory enzymes involved in the glycolytic pathway, oxPPP, phosphofructokinase regulation, pyruvate metabolism, and TCA that are either up- (in red) or down-regulated (in blue) in OsiR vs. Par cells.

## 4. Discussion

Resistance to osimertinib in EGFR-mutant NSCLC remains a significant clinical challenge. Our study provides an integrated view of metabolic remodeling in OsiR lung cancer cells, combining exometabolomic, transcriptomic, and proteomic evidence. The data converge on a consistent theme: resistant cells undergo a complex reprogramming that extends beyond the canonical Warburg phenotype and instead relies on alternative routes that sustain energy, biosynthesis, and redox balance under drug pressure.

OsiR cells reorient their metabolism towards glycolysis and this metabolic adjustment is crucial to maintain energy production and support biosynthetic processes under therapy-induced stress. The exometabolome analysis revealed high concentrations of acetate and significant increases in pyruvate, acetaldehyde, and lactate among other metabolites in the extracellular medium of OR cells. The significant increase in acetate and pyruvate suggested the activation of the pyruvate-acetaldehyde-acetate (PAA) pathway, a metabolic adaptation associated with resistance that has never been previously reported in lung cancer therapy resistance. This metabolic modality catalyze the NADP+-dependent oxidation of acetaldehyde to acetate, generating NADPH. Having outlined the overall picture, we will now delve into the specific metabolic pathways that drive resistance to osimertinib.

*Glycolysis, pentose phosphate pathway, and tricarboxylic acid switch in OsiR cells*

The interplay between glycolysis, the PPP, and the TCA cycle provides cancer cells with remarkable metabolic flexibility, enabling them to dynamically switch between them and reroute carbon flux as needed. Among these pathways, the PPP, which branches from glycolysis, plays a dual function: generating NADPH for redox homeostasis and reductive biosynthesis, and producing ribose-5-phosphate for nucleotide, vitamin and cofactor biosynthesis. The oxidative phase of the PPP, initiated by glucose-6-phosphate dehydrogenase (G6PDH), converts glucose-6-phosphate (G6P) into ribulose-5-phosphate while generating NADPH, particularly in cells with high NADPH demand. Following the ox-PPP, the non-oxidative phase (non-ox-PPP), driven by TKT and transaldolase (ALDO), interconverts sugar phosphates to support nucleotide and amino acid biosynthesis. Thus, the flexibility of the PPP allows cells to adjust to varying metabolic needs while maintaining balance between NADPH production and ribose-5-phosphate biosynthesis. This tight PPP integration with glycolysis and gluconeogenesis allows cancer cells to dynamically adapt to metabolic stress while balancing NADPG and ribose -5-phosphate production. Notably, its intermediates can reintegrate into glycolysis or feed into gluconeogenesis, highlighting the interconnected adaptability of carbon metabolism in cancer cells.

Pyruvate, mainly generated from glycolysis, sit as the hub of this network, linking glycolysis to the TCA cycle. Additional pyruvate sources, well established in biochemistry, include conversion of malate (via malic enzymes), alanine (via alanine transaminase), and other gluconeogenic amino acids. Once formed, pyruvate is a critical metabolic junction. Pyruvate can either remain cytosolic and be reduced to lactate by lactate dehydrogenase (LDH), as in the Warburg effect, or transported into the mitochondrial matrix to fuel the TCA cycle as acetyl-CoA through the PDH complex.

Uncontrolled proliferation and therapy resistance in cancer are often linked to the Warburg effect, characterized by increased glycolysis and lactate production even in the presence of oxygen (Diaz-Ruiz et al., 2011; F Marcucci and Rumio, 2021; Peng et al., 2021; Xu et al., 2005). In OsiR cells, elevated levels of pyruvate (**Figure 2A**) reflect this metabolic shift, indicating enhanced glycolytic flux and redirection of pyruvate towards lactate, consistent with a Warburg-like profile. However, the metabolic phenotype of OsiR cells is not limited to this classical pathway and appears to extend beyond the Warburg effect. Emerging evidence points to additional metabolic adaptations beyond classical glycolysis. Notably, the accumulation of acetate (**Figure 2B**) we detected in the extracellular milieu of OsiR cells, suggests diversion of pyruvate away from lactate and its processing through alternative pathways under osimertinib-induced stress. In this context, acetate may act as a byproduct but also as a carbon source for neighboring cancer cells, reinforcing metabolic symbiosis and carbon sharing within the tumor environment.

*Hexokinases and Glycolytic Commitment*

Among the key regulatory enzymes, **hexokinases (HKs)** are high-affinity enzymes physically associated with the outer mitochondrial membrane, that catalyze the first and irreversible step of glycolysis, phosphorylating glucose to G6P, thus committing it to downstream metabolism. Hexokinase expression has been linked to metabolic adaptation in resistance (DeWaal et al., 2018). In OsiR cells the housekeeping isoform HK1 is upregulated and ensures that glucose is efficiently trapped within the cell and is subsequently fluxed into glycolysis, providing a steady supply of G6P for both glycolysis and the PPP. In contrast, HK2, a HIF1 transcriptional target, commonly contributes to glycolytic rewiring under hypoxia (DeWaal et al., 2018), but it seems less prominent in OsiR resistance..

*oxPPP Enzymes and Drug Resistance*

Oxidative PPP enzymes, including G6PDH and H6PDH catalyze the first committed step of the ox-PPP. In OsiR cells, their decreased gene expression is accompanied by downregulation of PGLS, which is the mutase whose reaction product serves as a substrate for PGD, the second dehydrogenase in the PPP. As a result, ox-PPP activity is reduced in OsiR cells. This reprogramming preferentially redirects glucose toward glycolysis, reflecting the Warburg effect, rather than toward the ox-PPP. Although glycolysis yields less ATP per glucose molecule compared to OXPHOS, this adaptation eventually provides energy at a faster rate and supports survival under drug-induced stress. However, this shift compromises NADPH production, which is typically supplied by the ox-PPP. Despite this, resistant cells maintain aggressive phenotypes, including their migratory and invasive capabilities, implying the presence of compensatory metabolic mechanisms that support their thriving under stress. Notably, G6PDH, the rate-limiting enzyme of the PPP, is crucial in mediating cancer drug resistance, providing NADPH to buffer oxidative stress (Song et al., 2022) and its mutations reduce NADPH levels, increase ROS, and reduce metastatic potential (Aurora et al., 2022). However, the overall relevance of G6PDH and H6PDH in cancer and drug resistance is still incompletely understood (Jin and Zhou, 2019).

*Phosphofructokinase Regulation*

**Phosphofructokinase** (PFK1) catalyzes the first committed step of glycolysis, converting fructose 6 phosphate (F6P) to fructose 1,6 biphosphate (F1,6BP). PFK1 reaction is irreversible, and F1,6BP is exclusively used for glycolysis. This enzyme is tetrameric, existing in three isoforms: PFKM (muscle), PFKL (liver), and PFKP (platelet). The **upregulation of PFKM** and **downregulation of**

**PFKL** in OsiR cells suggests that the predominant PFK1 complex is composed almost exclusively of PFKM subunits. This configuration favors glycolysis and higher F1,6BP levels. PFK1 activity is further regulated by fructose-2,6-bisphosphate (F2,6BP), a potent activator of PFK1 that dictates glycolytic flux (Yi et al., 2019). The enhancement of the glycolytic flux in OsiR cells is also favored by the upregulation of **PFKFB4** and **PFKFB3**, that control the cytoplasmic F2,6BP levels and allosterically activate PFK1. PFKFB4, which is overexpressed in lung adenocarcinomas, promotes oncogenic phenotypes (Meng et al., 2021) by redirecting glucose flux to the non-ox-PPP (via TKT), supporting nucleotide, amino acid, and vitamin biosynthesis (Meng et al., 2021). PFKFB3, although moderately upregulated, primarily supports glycolysis by enhancing PFK1 activity. The differential roles of these isoenzymes allow cancer cells to balance energy production, redox homeostasis, and biosynthetic needs under drug-resistance conditions. Notably, PFKFB4 exhibits higher fructose-2,6-bisphosphatase activity compared to PFKFB3, facilitating F2,6BP degradation, thus redirecting glucose flux to the non-ox-PPP (Shen et al., 2020). Overall, these metabolic adaptations couple glycolysis with the non-ox-PPP, providing cancer cells with metabolic flexibility needed to survive under therapeutic pressure.

ENO2 upregulation in OsiR is relevant not only for its glycolytic function but also for its non-glycolytic roles described in the literature, including modulation of epithelial-mesenchymal transition (EMT) and drug sensitivity, contributing to therapy resistance (Ni et al., 2023)(Pujol et al., 2001). Similarly ALDOA, which is also upregulated, enhances glycolysis and channels intermediates into the non-ox-PPP, promoting chemoresistance (Li et al., 2020).

LC-HRMS analysis detected decreased levels of G3P in OsiR cells (**Figure 2C**), reflecting its shunting to the non-ox-PPP via TKT. TKT increases the pool of non ox-PPP metabolites essential for nucleotide and cofactor biosynthesis by converting (i) G3P + F6P into xylulose-5-phosphate + erythrose-4-phosphate, and (ii) G3P and sedoheptulose-7- phosphate into D-xylulose-5- phosphate + ribose-5- phosphate. Ribose-5- phosphate is a critical precursor in proliferative cells for the biosynthesis of purine and pyrimidine nucleotides, as well as for coenzyme forms of several vitamins, including NAD, NADP, FAD, and CoA (Jiang et al., 2014). While transcriptomics shows that TKT is equally expressed in Par and OsiR cells, Western blot evidenced TKT overexpression in OsiR cells (Maroni et al., 2025). This overexpression correlates with the observed increase in ribose-5-phosphate, D-ribulose 5-phosphate, and D-Xylulose 5-phosphate in OsiR ECMs (**Figure 2D**). Consequently enhanced glycolysis coupled with increased TKT activity sustains non-ox-PPP flux, promoting nucleotide biosynthesis (Li et al., 2020).
Elevated PPP activity is a common feature of cancer cells (Nagashio et al., 2019)(Cossu et al., 2020), and TKT overexpression has been linked to poor prognosis across multiple cancers, including lung cancer. As a key enzyme of the non-ox-PPP, TKT significantly impacts cancer progression and patient survival (Cao et al., 2021; Hao et al., 2022; Niu et al., 2022). In NSCLC, TKT knockdown inhibits cell proliferation and enhances sensitivity to gefitinib, counteracting the effects of TKT overexpression (Cao et al., 2021), highlighting TKT as a promising therapeutic target for advanced lung cancer treatment.

In OsiR cells, G3P is also converted to form **3-phosphoglycerate (3PG),** that accumulates in OsiR cells, as observed by LC-HRMS **(Figure 2E)**. Phosphoglycerate mutase 1 **(PGAM1)** reversibly converts 3-PG to 2-PG, a crucial step in glycolysis. Interestingly, 3PG inhibits 6-phosphogluconate dehydrogenase (PGD) in the ox-PPP, while 2-PG activates 3-phosphoglycerate dehydrogenase (PHGDH), which catalyzes conversion of 3-PG into 3-phosphohydroxypyruvate (3-PHP), marking the committed step in L-serine biosynthesis. Recent *in vitro* and *in vivo* studies show that increased 3-PG promotes cancer progression via PHGDH activity (K. Wang et al., 2023) thus highlighting its pro-tumorigenic function.

Lactate dehydrogenase converts L-lactate and $NAD^+$ to pyruvate and NADH. Isoenzymes LDHA and LDHB control T cell glycolysis and differentiation (Chen et al., 2023), while **LDHC** is abundant in cancer cells, which rely on aerobic glycolysis for their energy requirements (Tan et al., 2022). In OsiR cells, LDHC is upregulated and preferentially expressed over LDHA (Maroni et al., 2025). LDHC's high catalytic activity at acidic pH and elevated temperatures supports the Warburg effect (Tan et al., 2022), consistent with the high lactate concentration measured in the OsiR ECM (0.606±0.047 vs 0.429±0.021 nmol/cell in Par cells) (**Figure 2F** and **Table 2**).

Among **TCA-related genes,** Par and OsiR cells differ in aconitase (ACO1/IRP1), succinate dehydrogenase, succinyl-CoA ligase ADP-forming subunit β (SUCLA2), succinate-CoA ligase GDP-Forming subunit β (SUCLG2), isocitrate dehydrogenase (IDH), adenosine triphosphate citrate lyase (ACLY), and the PDH complex. OsiR cells exhibit upregulation of mitochondrial **ACO1/IRP1,** which is required for citrate/isocitrate interconversion in the TCA cycle, mitochondrial genome maintenance, and iron homeostasis (Mirhadi et al., 2023).

Transcriptomic analysis revealed upregulation of multiple SDH subunit geness (**SDHC**, SDHB , and SDHAF1), suggesting enhanced activity of mitochondrial Complex II (MCII) within both the TCA cycle and the ETC. Western blot further confirmed SDH overexpression in OsiR cells. SDH MCII is composed of four nuclear-encoded subunits and is the only enzymatic complex operating at the intersection of the TCA cycle and ETC, catalyzing the oxidation of succinate to fumarate while reducing ubiquinone to ubiquinol, thus transferring electrons from the TCA cycle to OXPHOS. Unlike other ETC complexes, MCII lacks mitochondrial-encoded subunits, explaining why defects in this complex are relatively rare (Q. Wang et al., 2023). Among the upregulated subunits, **SDHC** plays an important role in energy generation to maintain cellular growth (Q. Wang et al., 2023), **SDHB** encodes the iron-sulfur subunit that mediates electron transfer within MCII (Kita et al., 1990), and **SDHAF1**, a mitochondrial assembly factor, is indispensable for SDH assembly, despite not physically associating with the complex *in vivo*. The SDH complex also contributes to oxygen sensing by regulating succinate, an oxygen-sensor metabolite that stabilizes hypoxia-inducible factor 1 (HIF1) (Atallah et al., 2022).

Similarly, two genes encoding subunits of succinyl-CoA synthetase (SCS) show opposite transcriptional regulation in OsiR cells: **SUCLG2** is upregulated, while SUCLA2 is downregulated. SUCLG2 encodes the GTP-specific β-subunit of SCS, and has been implicated in mitochondrial dysfunction and the regulation of protein succinylation in lung adenocarcinoma (Hu et al., 2023). **SUCLA2** encodes the ATP-specific β-subunit of SCS, forming a heterodimer with the α-subunit. Interestingly, SUCLA2's role in cancer appears to extend beyond canonical TCA cycle metabolism: it contributes to redox balance and promotes survival of disseminated cancer cells during metastasis, independently of its enzymatic function within the TCA cycle (Boese et al., 2023).

**Isocitrate dehydrogenases** catalyze the oxidative decarboxylation of isocitrate to α-ketoglutarate (α-KG). These enzymes belong to two distinct subclasses, one of which utilizes $NADP^+$ as the electron acceptor. In OsiR cells, IDH-encoding genes are downregulated (**Figure 3**), including cytoplasmic/peroxisomal **IDH1** and mitochondrial **IDH2**. IDHs are central in intermediary metabolism and energy production and may interact closely with the PDH complex. In OsiR cells, the NAD$^+$-dependent non-catalytic subunit γ of IDH, **IDH3G**, which can be allosterically activated by citrate and/or ADP, is also downregulated, suggesting citrate is redirected toward the cytoplasm as substrate for **ACLY**. Supporting this, **SLC25A1**, encoding the mitochondrial citrate carrier (CIC), is similarly downregulated in OsiR cells.

**ACLY**, which catalyzes the ATP-dependent conversion of citrate and CoA into oxaloacetate (OAA) and acetyl-CoA for lipogenesis, is upregulated in OsiR cells. Relatedly, the gene encoding dehydrogenase/reductase member 2, **DHRS2**, a member of the short-chain dehydrogenases/reductases (SDR) family, shows remarkable upregulation. SDR enzymes are involved in reprogramming lipid metabolism and redox homeostasis to regulate proliferation, migration, invasion, and drug resistance in cancer cells (Li et al., 2021).

Pyruvate dehydrogenase serves as a central gatekeeper of glucose metabolism, directing pyruvate either towards the TCA cycle as acetyl-CoA or toward lactate production. Its activity is tightly regulated by pyruvate dehydrogenase kinase (PDK), which inhibits PDH through phosphorylation, and by a corresponding phosphatase that restores PDH function. Consequently, PDH inhibition or PDK overexpression shift metabolism towards glycolysis. In OsiR cells, both **PDK1** and PDHA1, which encodes the α-subunit of the E1 component of PDH and serves as the primary phosphorylation target of PDK1, are downregulated. Conversely, PDHB is upregulated, potentially subtly altering PDH transcriptional stoichiometry. Despite these transcriptional changes, Western blot analysis demonstrates significant PDH protein overexpression in OsiR cells (Maroni et al., 2025), specifically of PDHE1 subunit (**Figure 2I**), which is involved in an atypical decarboxylation step of pyruvate. Dysregulation of PDHA1 has been implicated in metabolic reprogramming, and variable PDHA/PDHB expression patterns are reported across cancer types and drug resistance models, reflecting diverse metabolic adaptative strategies (Zhang et al., 2023). In addition, PHDB has been reported to contribute to drug resistance in lung cancer by promoting EMT (Cevatemre et al., 2021).

The precise mechanisms underlying the metabolic adaptations of OsiR cells remain incompletely elucidated. Under normal conditions, pyruvate is transported into the mitochondrial matrix via the mitochondrial pyruvate carrier (MPC), where pyruvate carboxylase (PC) converts it into OAA. Oxalacetic acid not only fuels the TCA cycle but also serves as a precursor for numerous biosynthetic pathways (gluconeogenesis, nucleotide, amino acid synthesis). Many tumors downregulate MPC as a strategy to limit mitochondrial pyruvate oxidation, thus sustaining the Warburg phenotype, preserving cytosolic pyruvate for lactate production, and redirecting carbons into anabolic pathways (Furusawa et al., 2023). Notably, MPC1 has been shown to sensitize NSCLC cells to PARP inhibition, linking mitochondrial pyruvate transport to therapeutic vulnerability (Furusawa et al., 2023).

Beyond the TCA genes and associated proteins discussed above, the other TCA cycle enzymes are expressed at comparable levels in Par and OsiR cells. Nevertheless, mutations in ETC complexes (MCI, MCIII, MCIV and ATPase), together with functional assays (Maroni et al., 2025), reveal compromised ETC OXPHOS function and severely impaired mitochondrial ATP production in OsiR cells.

These observations suggest that mitochondria are repurposed from their canonical bioenergetic *modus operandi* to prioritize biosynthetic and anaplerotic processes rather than ATP production. Metabolites such as OAA and α-KG serve as building blocks for amino acid synthesis and provide alternative entry points for carbon derived from amino acids. This flexibility implies that when the TCA cycle is impaired by stress or mutations, it may no longer function primarily as an energy-generating engine (as previously thought); instead cells adapt by redirecting metabolism to maintain functionality by sustaining essential cellular functions (Ciccarone et al., 2018; Ciccarone and Ciriolo, 2024). A provocative hypothesis is that mitochondria can voluntarily "switch off" their bioenergetic capacity in response to specific signaling cues, *de novo* adopting the above-mentioned dedicated anaplerotic role to support survival and growth under metabolic stress.

*Pyruvate, acetaldehyde and acetate: an alternative metabolic route in OsiR cells*
Chromatographic analysis revealed marked differences in the exometabolome of Par and OsiR cells. In Par cells, lactate was the most abundant metabolite, followed by pyruvate and acetaldehyde, while acetate remained below the instrumental limit of detection (**Table 2**). In contrast, OsiR cells exhibited acetate as the predominant metabolite, followed by lactate, pyruvate, and acetaldehyde. The distinct composition of the ECM in Par versus and OsiR cells, together with their differing acid strengths (lactate $pK_a$ =3.1, and acetate $pK_a$=4.8), likely contributes to the increased ECM acidity observed in OsiR cells. Notably, the culture medium of OsiR cells appeared more yellow (Maroni et al., 2025), reflecting greater net acidification.

Elevated lactate production is a well-recognized feature of drug-resistant cancer cells, reflecting enhanced glycolysis associated with the Warburg effect. However, the unexpectedly high levels of pyruvate, acetaldehyde, and acetate in OsiR cells suggest that these cells selectively reroute glycolytic, TCA, and PPP intermediates to meet energy, redox, and biosynthetic demands. Given that hypoxia and acidosis (alone or in combination) trigger distinct gene expression and protein synthesis responses (Ciccarone et al., 2018), these metabolic shifts may have crucial, yet uncharacterized, implications for cell survival.

Proteins involved in **pyruvate** metabolism are frequently differentially expressed in cancer versus normal cells, and key glycolytic enzymes contribute directly to drug resistance (Fabrizio Marcucci and Rumio, 2021). Downregulation of mitochondrial oxidative metabolism (long-term Warburg effect) can help cancer cells escape cell death (Diaz-Ruiz et al., 2011). Additionally, in the presence of glucose, cancer cells can transiently suppress respiration, a phenomenon known as the **"Crabtree effect"** (Diaz-Ruiz et al., 2011), although its precise mechanism remains unclear. Diaz-Ruiz and colleagues noted parallels between glucose-induced repression of oxidative metabolism in *S. cerevisiae*, which favors fermentation, and the "aerobic glycolysis" observed in tumor cells (Diaz-Ruiz et al., 2011), underscoring conserved metabolic strategies across species. This metabolic reprogramming is particularly evident in tumor-specific isoforms of glycolytic enzymes, which facilitate adaptation to therapeutic stress (Fabrizio Marcucci and Rumio, 2021). When mitochondrial pyruvate transport and acetyl-CoA formation are reduced, cytosolic pyruvate levels rise (Herzig et al., 2012).

Cytosolic pyruvate levels are tightly controlled in both yeast and cancer cells. In *S. cerevisiae*, pyruvate accumulates in the cytosol during fermentation while its entry into the TCA cycle is limited, a process controlled by PDH kinase and phosphatase activities, as well as transcriptional regulation of the E3 subunit (lipoamide dehydrogenase). Similarly, in cancer cells, downregulation of TCA activity or impaired respiratory chain function restricts pyruvate utilization in mitochondria, leading to cytosolic accumulation. This pyruvate is the rerouted into alternative metabolic pathways that support the anaplerotic function of mitochondria (Ciccarone et al., 2018; Ciccarone and Ciriolo, 2024), sustaining tumor growth and drug resistance. Moreover, tumor-specific isoforms of glycolytic enzymes amplify these adaptive responses, reinforcing cellular survival under metabolic stress and therapeutic pressure (Fabrizio Marcucci and Rumio, 2021).

The role of **acetate** metabolism in cancer cells was first recognized in 1957, when Weinhouse et al. demonstrated that acetate can function as a respiratory substrate (MEDES et al., 1957)(Schug et al., 2016). Acetate serves as a versatile carbon source for lipid synthesis and energy production, especially under metabolic stress or hypoxic conditions (Wang et al., 2024). However, its precise origin, contribution to tumor metabolism, and potential role in immune evasion remain unclear (Wang et al., 2024), deserving further investigation.

In OsiR cells, acetate emerges as the predominant short-chain fatty acid dominating the ECM, suggesting that resistant cells preferentially rely on acetate to sustain biosynthesis of essential cellular components under Osimertinib treatment. Integration of metabolomic, transcriptomic, and protein-level analysis revealed a coherent metabolic axis underlying this metabolic reprogramming (Maroni et al., 2025). Acetate was detected exclusively in the OsiR ECM (**Figure 2B** and **Table 2**), accompanied by significantly higher acetaldehyde levels relative to Par cells (**Figure 2G** and **Table 2**). Notably, acetaldehyde concentrations were three orders of magnitude lower than acetate, likely reflecting its higher volatility, limited membrane diffusion, and rapid intracellular detoxification. If generated through imbalances in substrate and cofactor availability affecting PDH activity, as previously suggested by Liu et al. (Liu et al., 2018), acetaldehyde subcellular distribution would be dictated by PDH localization within the inner mitochondrial membrane, diffusing into both the mitochondrial intermembrane space and the matrix (Pelley, 2012).

However, before significant amounts of acetaldehyde can exit the cell, it is efficiently eliminated (given its high toxicity) by oxidation to ethanol (via ADH/CYP2E1/ in liver) or acetate (via NADPH-dependent ALDH enzymes) (Colombaioni et al., 2017). Consistent with this, ethanol was undetectable in the exometabolomes of both Par and OsiR cells.

Given the µM-range intracellular acetaldehyde concentrations reported by Liu et al. (Liu et al., 2018) the upregulation of glycolytic pathways and ALDH activity in OsiR cells likely drive the observed high µM-level accumulation of extracellular acetate.

The abundant acetate, coupled with increased PDH expression in OsiR cells, indicates that pyruvate is routed to form acetate through a defined acetate-producing pathway (**Scheme 1**), complementing enhanced glycolysis. This observation is consistent with previous reports in multiple human cancer cell lines displaying high acetate concentrations (Liu et al., 2018), and highlight acetate as a central adaptive metabolite in therapy-resistant metabolism.

Importantly, our findings position acetate not merely as a byproduct but as a central adaptive metabolite, forming a dedicated metabolic axis that supports biosynthesis and redox balance in Osimertinib-resistant cells.

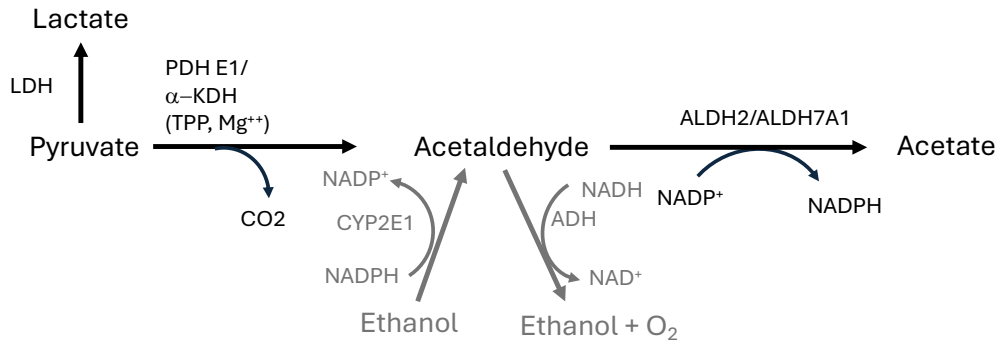

**Scheme 1**. Formation of acetate in OsiR through the pyruvate-acetaldehyde-acetate (PAA) pathway.

To date, 19 human ALDH genes have been identified, encoding cytosolic, mitochondrial, and microsomal enzymes expressed across multiple tissues (Nguyen et al., 2025; Tahiliani et al., 2025; Xia et al., 2023). ALDHs primarily catalyze the oxidizing of acetaldehyde to acetate (**Scheme 1**) (Diaz-Ruiz et al., 2011; Enomoto et al., 1991; He et al., 2015; Marchitti et al., 2010; Pramono et al., 2020; Yang et al., 2018), with mitochondrial class II aldehyde dehydrogenase (ALDH2) acting as the principal isoform for acetaldehyde clearance (Marchitti et al., 2008).

ALDHs are closely associated with tumor cell proliferation, migration, invasion, and therapy resistance, with distinct isoform-specific expression patterns across tumors (Dinavahi et al., 2019; He et al., 2015; Marchitti et al., 2008; Yang et al., 2018). For instance, 97.7% of cisplatin-resistant A549/DDP lung adenocarcinoma cells are ALDH-positive, and pharmacological inhibition of ALDH with DEAB significantly reduces resistance (He et al., 2015). Similarly, ALDH1A1 has been implicated in gefitinib resistance of lung cancer HCC-827/GR cells (Yang et al., 2018).

In OsiR cells, several ALDH isoforms are differentially expressed (**Figure 3**), including ALDH2, ALDH7A1, ALDH5A1, ALDH3B1, ALDH6A1, and ALDH3A2, reflecting a substantial rearrangement of acetaldehyde metabolism. ALDH2 and ALDH7A1 are upregulated at both mRNA and protein levels in OsiR cells (Maroni et al., 2025). ALDH2, a mitochondrial matrix enzyme constitutively expressed across tissues (Marchitti et al., 2008), exhibits high-substrate specificity for acetaldehyde ($K_m$ = 0.2-0.59 µM), outperforming ALDH1, and contributes to oxidative stress protection (Wang et al., 2009) (Klyosov et al., 1996; OHTA et al., 2004). ALDH2 is the primary enzyme involved in the oxidation of acetaldehyde ($K_m < 1$ µM) during ethanol metabolism (Klyosov et al., 1996). Mitochondrial ALDH2 upregulation is associated with chemoresistance, as seen in Paclitaxel-resistant NSCLC, and can be targeted using Disulfiram plus copper to reverse microtubule inhibitor resistance by suppressing ALDH2 expression (Wang et al., 2022). Overexpression of ALDH2 or ALDH1A2 similarly enhances proliferation, clonogenic potential, and chemoresistance to 4-hydroperoxycyclophosphamide and doxorubicin, in leukemia (K562 cells) or NSCLC (H1299 cells) models, respectively (Moreb et al., 2012).

ALDH7A1, localized to the cytosol, nucleus and mitochondria in a tissue-specific manner (at least in murine models) catalyzes $NAD(P)^+$-dependent oxidation of aldehydes and participates in the pipecolic acid pathway of lysine catabolism. It protects cells against hyperosmotic stress and attenuates reactive-aldehyde- and oxidative-stress-induced cytotoxicity (Brocker et al., 2011). Elevated lysine and pipecolic acid levels in OsiR cells are consistent with ALDH7A1 upregulation (data not shown). Kinetic characterization of mammalian ALDH7A1 activity towards other substrates remains limited (Marchitti et al., 2008).

Conversely, ALDH isoforms downregulated in OsiR cells, including ALDH5A1, ALDH3B1, ALDH6A1 and ALDH3A2, are less relevant to acetaldehyde oxidation. ALDH5A1 is a mitochondrial $NAD^+$-dependent succinic semialdehyde dehydrogenase, while ALDH3B1 is a cytosolic $NAD^+/NADP^+$-dependent enzyme that, similarly to ALDH3A2, is catalytically active toward C16 aldehydes derived from lipid peroxidation, suggesting a potential role against oxidative stress (Kitamura et al., 2013; Marchitti et al., 2010). ALDH3B1 displays low affinity for acetaldehyde ($K_m$=23.3 mM), and high affinity for hexanal ($K_m$=62 µM), octanal ($K_m$=8 µM), 4-hydroxy-2-nonenal ($K_m$=52 µM), and benzaldehyde ($K_m$=46 µM) (Marchitti et al., 2010).

ALDH inhibitors are under active investigation for cancer therapy to target chemoresistance (Dinavahi et al., 2019; Wang et al., 2018), yet the broader contributions of ALDHs to cancer metabolism remain incompletely understood.

*Why is the PAA metabolic reprogramming activated under life-threatening stress like chemotherapy?* The biological significance of the PAA pathway (**Scheme 1**) has never been reported. Liu et al. noted they "have by no means defined the general biological role of these reactions" (Liu et al., 2018). We hypothesize that high extracellular acetate concentration observed in OsiR cells results from an alternative ATP- generating route taht prioritize speed over yield, providing glycolysis-derived ATP and reducing power (NADPH) under metabolic stress. In this context, ALDH2 and ALDH7A1 have a specific role in NADPH production (**Scheme 1**), purposely switched on to support cell survival during drug resistance. This interpretation aligns with our previous findings in OsiR cells including: (i) upregulation of glycolytic-pathway proteins (Warburg effect), (ii) downregulation of the ox-PPP (thus limiting PPP-dependent NADPH production), (iii) enhanced lipid biosynthesis, and (iv) impaired mitochondrial OXPHOS (Maroni et al., 2025).

Bose et al. also reported that under conditions of mitochondrial dysfunction, hypoxia, limited metabolic resources, or within the tumor microenvironment, cells shift toward acetate as a major energy source, bypassing the canonical conversion of pyruvate to acetyl-CoA (Bose et al., 2019). Beyond serving as a carbon source and supporting immune evasion (Wang et al., 2024), the PAA pathway emerges a critical in sustaining cell survival under stress.

Acetyl-CoA can also be generated in mitochondria from pyruvate via PDH and in the cytosol from citrate via ACLY, which, as reported above, is upregulated in OsiR cells. Both processes are coupled to the TCA cycle. Additionally, acetyl-CoA synthetases **ACSS1** and **ACSS2** catalyze the conversion of endogenous acetate to acetyl-CoA in the mitochondria and cytosol, respectively, providing essential substrates for lipogenesis (Wang et al., 2024). Notably, ACSS1 expression is significantly higher in OsiR cells than in Par cells (**Figure 2J**), whereas ACSS2 has been shown to sustain acetyl-CoA pools and support cell proliferation under metabolically limiting conditions, including mitochondrial dysfunction or ATP citrate lyase (ACLY) deficiency (Comerford et al., 2014; Schug et al., 2016). These findings highlight the importance of ACSS enzymes in maintaining cellular metabolism under stress. Furthermore, the downregulation of isocitrate dehydrogenases IDH1 and IDH2 in OsiR cells suggests a reorientation of citrate metabolism towards reinforcing the acetyl-CoA pool.

NADPH is primarily consumed in anabolic reactions, such as fatty acid biosynthesis, while NADH, when mitochondrial function is intact or partially preserved, predominantly fuels catabolic pathways and energy production (Ju et al., 2020). Cellular NADH/NADPH balance is regulated by nicotinamide-nucleotide transhydrogenase (NNT), a membrane-bound enzyme that facilitates the reduction of $NADP^+$ by NADH through proton translocation. In mitochondria, NNT translocates protons from the intermembrane space to the matrix, while in the endoplasmic reticulum (ER), it moves protons from the ER lumen to the cytosol. Through this activity, NNT impacts redox status, biosynthesis, detoxification (via glutathione/thioredoxin systems), and apoptosis, and has been implicated in cancer cell proliferation (Liu et al., 2018; Ward et al., 2020). However, the role of NNT in NSCLC remains debated (He et al., 2020; Ward et al., 2020). In both Par and OsiR cells, NNT transcription and protein levels remain unchanged (data not shown), likely because NADPH/NADH ratios are similar, despite these cofactors being produced through different pathways, ox-PPP in Par cells, and the PAA pathway in OsiR cells.

## 4. Conclusions

Pathway enrichment analysis of the exometabolome revealed upregulation of glycolysis/gluconeogenesis, pyruvate metabolism, nicotinate and nicotinamide metabolism, and several amino acid pathways. By integrating these results with targeted metabolite quantitation, transcriptomic data, and protein validation by Western blot, we uncovered a coordinated cellular strategy to preserve viability under therapeutic pressure through the reprogramming of energy and redox metabolism. Specifically, glycolysis appeared decoupled from downstream oxidative metabolism, with enhanced activity evidenced by the accumulation of pyruvate, lactate, acetaldehyde, and acetate.
In Par cells, lactate dominated extracellular metabolite profile, followed by pyruvate and acetate (the latter below the limit of detection). Impressively, in OsiR cells, acetate emerged as the predominant contributor to extracellular acidity, followed by lactate and pyruvate. These findings indicate that resistant cells may rely more heavily on glycolysis (Warburg effect), supported by overexpression of key glycolytic enzymes and activation of alternative metabolic pathways that sustain survival under osimertinib treatment. The markedly elevated acetate levels, alongside other unique metabolic features in OsiR cells, suggest a specific resistance-associated metabolic switch.

Our recent transcriptomics, mitochondrial genomic, and imaging analyses of OsiR cells (Maroni et al., 2025) revealed profound alterations in mitochondrial function and morphology, including ETC impairment ETC, mtDNA mutations, and downregulation of several mitochondrial genes, all consistent with compromised OXPHOS.

We propose that the overproduction of acetate, alongside elevated pyruvate, reflects activation of an unconventional **pyruvate-acetaldehyde-acetate pathway**. This pathway likely serves as a primary source of NADPH, supporting lipid biosynthesis and antioxidant defense in resistant cells. Consistently, ALDH2 and ALDH7A1, enzymes catalyzing $NADP^+$-dependent oxidation of acetaldehyde to acetate, thus generating NADPH, were upregulated in OsiR cells.

Interestingly, while enzymes of the oxidative PPP (G6PDH, H6PDH, PGLS) were downregulated in OsiR cells, non-oxidative PPP enzymes were are expressed at comparable levels to Par cells. This suggests that both NADPH and the non-ox-PPP metabolites required for *de novo* lipid biosynthesis are replenished via alternative routes: in Par cells, NADPH mainly derives from the ox-PPP, while in OsiR cells, NADPH is generated via the PAA pathway, and metabolites are produced from glycolytic intermediates (F6P and G3P) via the "shunt" enzyme TKT. Both NADPH and acetate are crucial for acetyl-CoA production and *de novo* lipid biosynthesis.

The TCA in OsiR cells appears to serve primarily an anaplerotic role, providing precursors for amino acids, nucleotides, vitamin, and cofactors. Western blot analyses confirmed increased expression of PDH and SDH, key mitochondrial enzymes, supporting this function.

Together, these data reveal a comprehensive metabolic adaptation: despite mitochondrial bioenergetic capacity, OsiR mitochondria continue to sustain essential biosynthetic processes, while stress response pathways are upregulated, ensuring cell survival under therapeutic pressure.

**Competing Interest Statement**
The authors have declared no competing interest.

**FUNDER INFORMATION DECLARED**

AIRC Investigator Grant 2021 (ID 25734), PNNR MCNT1-2023-12377671 Grant, ELMO Fondazione Pisana per la Scienza FPS Grant 2024, and private donations from the Gheraldeschi and Pecoraro families to EL; European Union Next Generation EU (Mission 4 Component 2, project "Strengthening BBMRL.it") to AC; NUS A-0001263-00-00 to AA.

**Figures and tables with legends.**

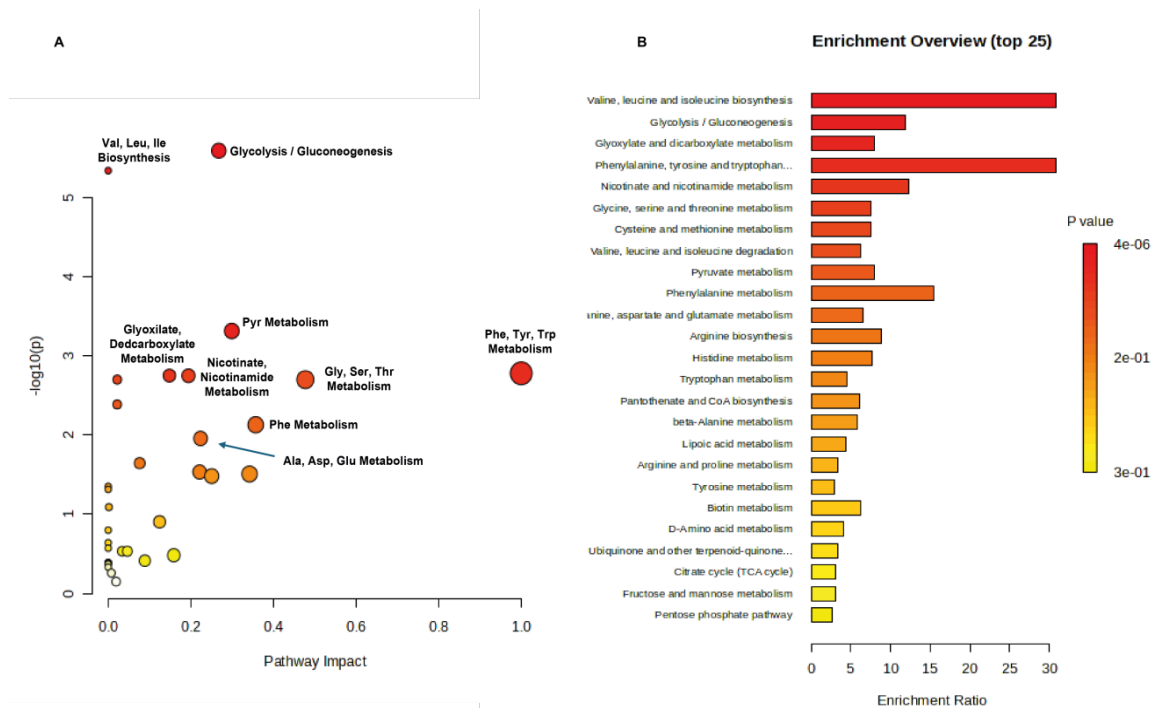

**Figure 1. (A)** Pathway analysis of significantly altered exometabolites in OsiR *vs* Par cells determined by LC-HRMS, LC-DAD and GC-MS. Pathways were mapped via the KEGG database and displayed according to pathway impact (dot size) and adjusted *p*-value (color intensity; darker = more significant). **(B)** Metabolite set enrichment analysis confirmed the most impacted pathways in OsiR vs Par cells. Statistical and topological analyses were performed using MetaboAnalyst 6.0(Pang et al., 2024b, 2024a). Pathway impact represents a combination of pathway enrichment and topology analysis (centrality). Circle size indicates pathway impact, while color intensity represents statistical significance (more intense red color corresponds to lower *p*-values, and higher significance).

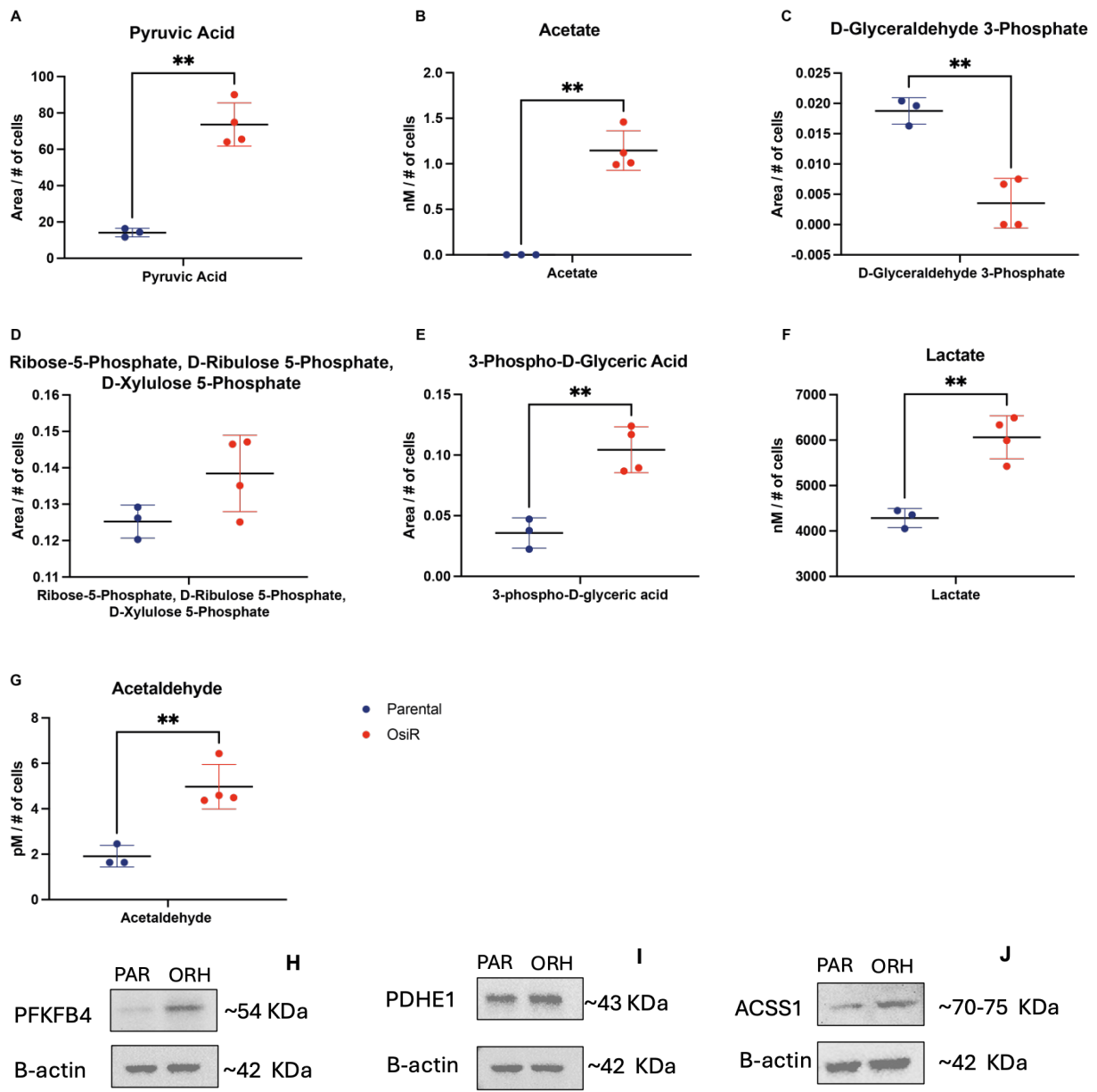

**Figure 2**. **(A-G)** Bar plots showing the peak areas (concentrations) of pyruvate, acetate, lactate, acetaldehyde, and other glycolysis/gluconeogenesis-associated metabolites measured by LC-HRMS, HPLC-DAD and HS-GCMS in Par (blue dots) and OsiR (red dots) ECMs. Statistical significance: $p < 0.001$ (**); $p<0.05$ (*). **(H-J)** Western blot analyses of Par and OsiR cells. Protein lysates were immunoblotted with an anti-PFKFB4, anti-PDHE1 and anti-ACSS1 antibody. Loading was assessed with an anti-β-actin antibody. Expected size is indicated in kDa.

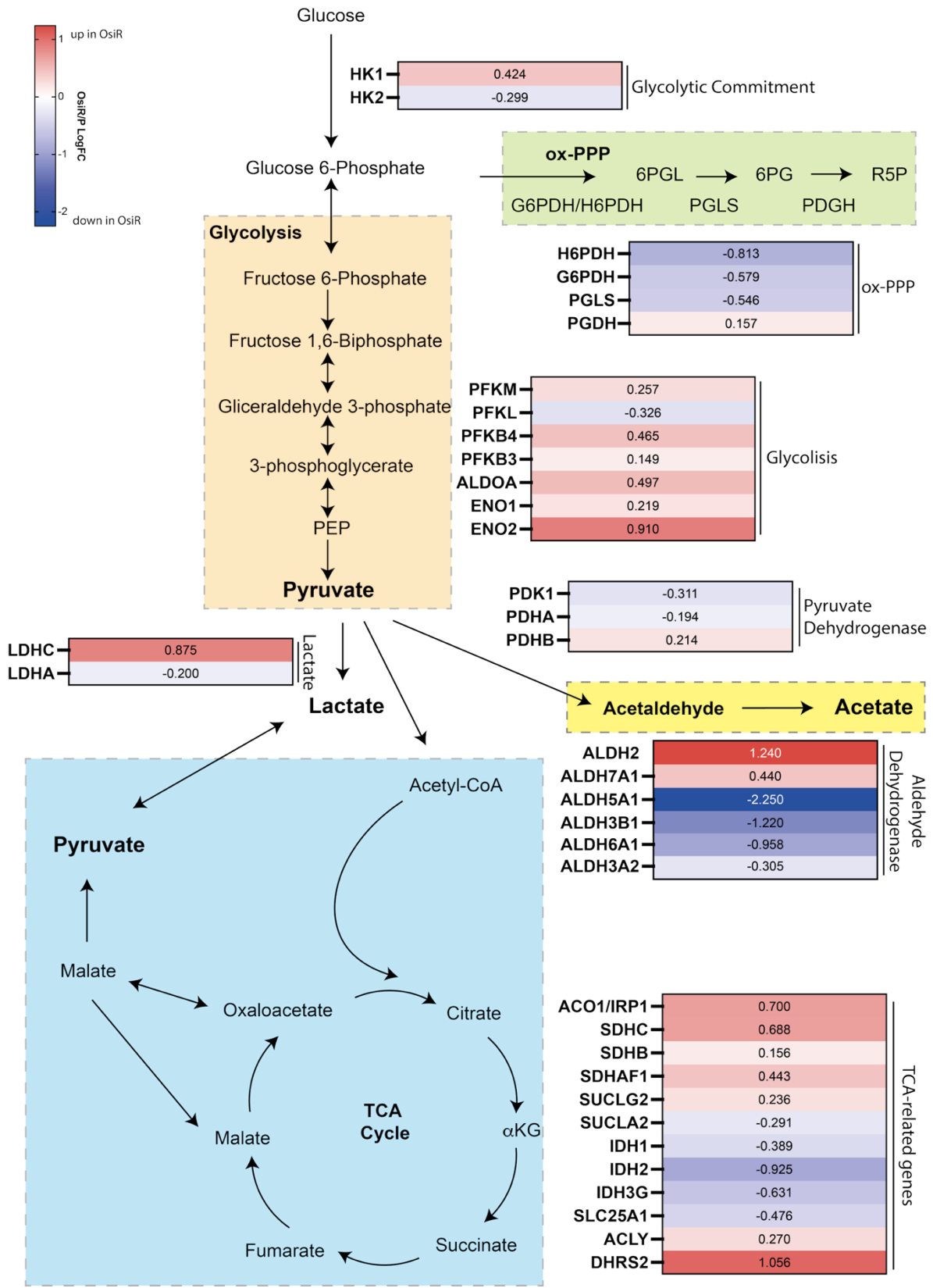

**Figure 3**. Heatmap of key regulatory enzymes involved in the glycolytic pathway, oxPPP, phosphofructokinase regulation, pyruvate metabolism, and TCA that are either up- (in red) or down-regulated (in blue) in OsiR vs. Par cells.

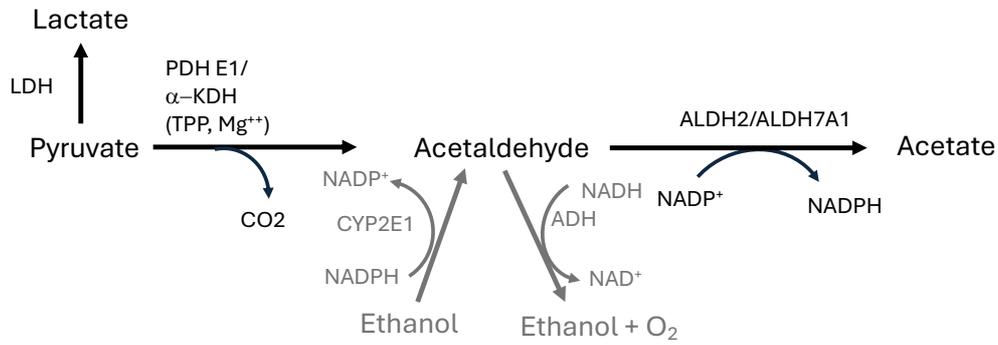

**Scheme 1**. Formation of acetate in OsiR through the pyruvate-acetaldehyde-acetate (PAA) pathway.

Table 1. Top 10 significantly dysregulated metabolic pathways in OsiR cells.

|  | Total | Expected | Hits | Raw $p$ | -log10($p$) | Holm adjust | FDR | Impact |
|---|---|---|---|---|---|---|---|---|
| **Glycolysis / Gluconeogenesis** | 26 | 0.44571 | 6 | 2.5748E-06 | 5.5893 | 0.00020598 | 0.00018353 | 0.2677 |
| **Valine, leucine and isoleucine biosynthesis** | 8 | 0.13714 | 4 | 4.5883E-06 | 5.3383 | 0.00036248 | 0.00018353 | 0 |
| **Pyruvate metabolism** | 23 | 0.39429 | 4 | 0.00048638 | 3.313 | 0.037937 | 0.01297 | 0.29918 |
| **Phenylalanine, tyrosine and tryptophan biosynthesis** | 4 | 0.068571 | 2 | 0.0016632 | 2.779 | 0.12807 | 0.019978 | 1 |
| **Glyoxylate and dicarboxylate metabolism** | 32 | 0.54857 | 4 | 0.0017764 | 2.7505 | 0.13501 | 0.019978 | 0.14815 |
| **Nicotinate and nicotinamide metabolism** | 15 | 0.25714 | 3 | 0.0017838 | 2.7487 | 0.13501 | 0.019978 | 0.1943 |
| **Cysteine and methionine metabolism** | 33 | 0.56571 | 4 | 0.0019978 | 2.6995 | 0.14783 | 0.019978 | 0.02184 |
| **Glycine, serine and threonine metabolism** | 33 | 0.56571 | 4 | 0.0019978 | 2.6995 | 0.14783 | 0.019978 | 0.47737 |
| **Valine, leucine and isoleucine degradation** | 40 | 0.68571 | 4 | 0.0041085 | 2.3863 | 0.29581 | 0.03652 | 0.02168 |
| **Phenylalanine metabolism** | 8 | 0.13714 | 2 | 0.007439 | 2.1285 | 0.52817 | 0.059512 | 0.35714 |

Table 2. Concentration levels of pyruvate, lactate, acetate, and acetaldehyde in the ECM of Par and OsiR cell cultures. Measurements were performed on three biological replicates and normalized to cell numbers. LOD= limit of detection.

| Concentration | Method | Unit | mean ± SD (Par) | mean ± SD (OsiR) |
|---|---|---|---|---|
| Pyruvate | (LC-HRMS) | (nmol/cell) | 0.022±0.004 | 0.113± 0.018 |
| Lactate | (LC-DAD) | (nmol/cell) | 0.429±0.021 | 0.606± 0.047 |
| Acetate | (HS-GCMS) | (nmol/cell) | <LOD | 1.144± 0.219 |
| Acetaldehyde | (HS-GCMS) | (pmol/cell) | 0.0019±0.0005 | 0.0050± 0.0009 |